\documentclass[twocolumn,aps,prb,amsmath,floatfix]{revtex4}
\usepackage{graphicx}
\begin{document}
\title{Coulomb correlations in the honeycomb lattice: 
role of translation symmetry}
\author{Ansgar Liebsch$^1$ and Wei Wu$^2$}
\affiliation{$^1$Peter Gr\"unberg Institut, Forschungszentrum J\"ulich, 
             52425 J\"ulich, Germany \\
             $^2$D\'epartement de Physique and RQMP, Universit\'e de Sherbrooke, 
             Sherbrooke, Qu\'ebec, J1K 2R1, Canada
 }
\begin{abstract} 
The effect of Coulomb correlations in the half-filled Hubbard model of the honeycomb 
lattice is studied within the dynamical cluster approximation (DCA) combined with 
exact diagonalization (ED) and continuous-time quantum Monte Carlo (QMC). 
The important difference between this approach and the 
previously employed cluster dynamical mean field theory (CDMFT) is that DCA
preserves the translation symmetry of the system, while CDMFT violates 
this symmetry. As the Dirac cones of the honeycomb lattice are the consequence of 
perfect long-range order, DCA yields semi-metallic behavior at small onsite 
Coulomb interactions $U$, whereas CDMFT gives rise to a spurious excitation gap 
even for very small $U$. This basic difference between the two cluster 
approaches is found regardless of whether ED or QMC is used 
as the impurity solver. At larger values of $U$, the lack of translation 
symmetry becomes less important, so that the CDMFT reveals a Mott gap, in 
qualitative agreement with large-scale QMC calculations. In contrast, the  
semi-metallic phase obtained in DCA persists even at $U$ values where CDMFT 
and large-scale QMC consistently show Mott insulating behavior.            
\end{abstract}
\maketitle

\section{Introduction}

The possible existence of a spin-liquid phase on the honeycomb lattice 
has recently attracted considerable attention. Meng {\it et al.}\cite{meng} 
investigated the Hubbard model for this system at half-filling, using large-scale 
quantum Monte Carlo (QMC) calculations for clusters containing up to 648 sites. 
Careful finite-size extrapolations indicated semi-metallic behavior for onsite 
Coulomb interactions in the range $U\le 3.5 t$ ($t$ is the nearest neighbor
hopping) and an antiferromagnetic insulator for $U\ge 4.3t$. The intermediate 
range $3.5t\le U \le 4.3t$ then corresponds to a Mott phase without long-range
order, the hallmark of a spin liquid. These findings were, however, disputed
by Sorella {\it et al.}\cite{sorella} who performed similar QMC calculations 
for even larger clusters including up to 2592 sites. The new results showed a
considerably reduced spin-liquid phase, confined at most to the narrow window 
$3.8t \le U\le 3.9t$.              

The effect of nonlocal Coulomb correlations on the honeycomb lattice was also
studied within the cluster extension of dynamical mean field theory \cite{cdmft} 
(CDMFT). Wu {\it et al.}\cite{wu} used continuous-time quantum Monte Carlo
\cite{rubtsov} (CTQMC), whereas Liebsch \cite{PRB2011} employed a 
multi-orbital-multi-site extension \cite{perroni,tong} of finite-temperature exact 
diagonalization \cite{ed} (ED) as impurity solver. Despite the fact that in 
the ED CDMFT calculations it was possible to include only a relatively small
bath (six bath levels per six-site unit cell), the cluster self-energy components
were found to be in nearly quantitative agreement with the CTQMC CDMFT results.
(For a detailed comparison see Fig.~25 of Ref.~\onlinecite{EDR}.) In particular, 
for $U\approx 5t$ both schemes revealed a Mott phase, with an excitation gap 
$\Delta\approx0.6t$, in close agreement with the one found by Meng {\it et al.}
\cite{meng} With decreasing $U$, the CTQMC results at temperatures $T\ge 0.05t$
indicated the closing of the Mott gap near $U=3.8t$, \cite{wu} while the ED 
results at lower temperature $T=0.005t$ revealed a weak insulating contribution 
to the self-energy at the Dirac points at arbitrarily low $U$.\cite{PRB2011} 
For $U\le3t$ the small gap associated with this self-energy was, however, 
difficult to resolve in the spectral distributions due to the temperature 
rounding of the gap edges.  
  
Analogous ED CDMFT calculations (also for six bath levels) were carried out by 
He and Lu \cite{helu} at a considerably lower effective temperature ($T=10^{-5}t$). 
The excitation gap in this case was found to extend to $U\rightarrow0$. 
On the basis of these results the authors concluded that the spin-liquid phase 
of the honeycomb lattice at half-filling exists from $U=0$ up to the onset of 
the antiferromagnetic phase near $U=4.5t$.

Closely related to these works are two calculations based on the variational 
cluster approximation \cite{VCA} (VCA) by Yu {\it et al.} \cite{yu} and 
Seki and Ohta.\cite{seki} In both cases, ED was used as impurity solver, with 
six bath levels as in Refs.~\onlinecite{PRB2011} and \onlinecite{helu}. 
Whereas Yu {\it et al.}~identified a spin-liquid phase in the range 
$U\approx 3t - 4t$ and semi-metallic behavior at smaller values of $U$,  
Seki and Ohta obtained a similar insulating contribution to the self-energy 
at the Dirac points as in Ref.~\onlinecite{PRB2011} and concluded that the 
Mott gap persists down to arbitrarily small values of $U$.  

Most recently, Hassan and S\'en\'echal \cite{hassan} performed ED calculations 
for the honeycomb lattice within VCA, CDMFT and the cluster dynamical impurity 
approximation \cite{CDIA} (CDIA).
They argued that a bath consisting only of six levels is insufficient and leads 
to the erroneous conclusion that the system is gapped for all nonzero values of 
the onsite Coulomb interaction $U$. 

In this context it is also important to recall the results of functional 
renormalization group (FRG) calculations \cite{frg} for the honeycomb lattice 
which reveal a stable semi-metallic phase below about  $U\approx 3.8 t$.
 
In view of these contradictory results it is evident that the possible existence 
and extent of the semi-metallic phase of the honeycomb lattice are difficult to 
determine within present non-local many-body techniques. In particular, it is not 
clear which assumptions and approximations give rise to certain consequences: 
the size of the correlated cluster, the size and symmetry of the bath in ED, 
the accessible temperature range, the accuracy of spectral functions at very low 
energies, etc.  Naturally, these uncertainties also affect the identification 
of the elusive spin-liquid phase.

The purpose of this work is to shed light on some of these issues by comparing 
new results derived within the dynamical cluster approximation \cite{DCA} (DCA)
with previous ones obtained within CDMFT.\cite{wu,PRB2011} As impurity solver 
we use finite-temperature ED as well as CTQMC. The nearly quantitative agreement 
between the ED and CTQMC self-energies, within DCA as well as CDMFT, demonstrates 
that the intrinsic limitations of these impurity solvers are not the cause of the 
discrepancies between the various results cited above. 

Instead we show here that, in the special case of the honeycomb lattice, it is
of crucial importance to preserve the translational invariance of the system.
Obviously, any deviation from bulk symmetry opens a gap at the Dirac points.
Recall, for instance, the single-particle gaps obtained for graphene ribbons.   
Thus, the semi-metallic and spin liquid phases can only be studied properly by 
using many-body methods that do not violate translation symmetry. 
This argument disqualifies CDMFT which is well-known to yield a self-energy 
that is not translationally invariant.\cite{DCA,biroli} The self-energy components
in this scheme account for correlations within the unit cell, but not between cells.
We therefore believe that all CDMFT calculations performed until now for the 
honeycomb lattice should exhibit, at low $U$ and low $T$, an excitation gap which 
is an artifact caused by the lack of translation symmetry of the self-energy. 
Although this gap is related to the presence of the local Coulomb interaction, 
it is not a true Mott gap but merely the consequence of the intrinsic limitation 
of the cluster approach. As a result, CDMFT and other schemes that do not preserve 
translation invariance are not suitable for the identification of a spin-liquid 
phase on the honeycomb lattice.

The comparison of the CDMFT self-energy with analogous results derived within DCA, 
for ED as well as CTQMC, underlines this point. In DCA, the self-energy is by 
construction translationally invariant, so that the electronic structure at low 
$U$ is semi-metallic, in agreement with the predictions based on large-scale QMC 
and FRG calculations.\cite{meng,sorella,frg} The spurious tail of the excitation 
gap at small $U$ and low $T$ that is seen in CDMFT is absent in DCA. 

As will be shown below, in the case of the honeycomb lattice, the DCA condition 
that ensures translation symmetry is too rigid for the description of 
correlations within the unit cell. As a result, the semi-metallic phase is still 
stable near $U=5t-6t$ where CDMFT and large-scale QMC calculations already find Mott 
insulating behavior. Thus, CDMFT and DCA may be viewed as complementary cluster 
schemes: DCA is preferable at low $U$ since it maintains the long-range order that 
is crucial for the Dirac cones, whereas CDMFT yields a more realistic description 
of short-range correlations in the Mott phase when the absence of translation 
symmetry plays a minor role.

We also note here that the gap tail obtained in CDMFT at small $U$ is not related 
to the finite size and symmetry properties of the bath used in ED.  
On the contrary, in the special case of the honeycomb lattice, a rather small 
bath containing only six levels is sufficient for the description of short-range 
correlations within the six-site unit cell. The reason is that, because of the 
semi-metallic properties of the system, the projection of the bath Green's function 
on a finite-cluster is not affected by the usual low-energy disparities that arise 
in the case of correlated metals.

The outline of this paper is as follows: In Section II we discuss the 
application of DCA and CDMFT to the honeycomb lattice and point out the 
key difference between the self-energies obtained within these schemes.     
Section III presents the main ingredients of the ED impurity solver for
both DCA and CDMFT. Section IV provides the discussion of the results obtained 
within ED DCA, and the comparison with analogous CTQMC DCA results.
The summary is presented in section V. Throughout this work only paramagnetic
phases are discussed.

\section{DCA vs. CDMFT for the Honeycomb Lattice}

To describe Coulomb correlations in the honeycomb lattice we consider the 
single-band Hubbard Hamiltonian 
\begin{equation}
H=-t \sum_{\langle ij\rangle\sigma} ( c^+_{i\sigma} c_{j\sigma} + {\rm H.c.}) 
                     + U \sum_i n_{i\uparrow} n_{i\downarrow} ,
\end{equation}
where $t$ is the nearest neighbor hopping term and $U$ the on-site Coulomb energy.
Throughout this paper $t=1$ defines the energy scale. The non-interacting band
dispersion is given by:    
$\epsilon({\bf k})= \pm t\vert 1+e^{ik_x\sqrt{3}}+e^{i(k_x\sqrt{3}+k_y3)/2}\vert$.
The nearest neighbor spacing is assumed to be $a=1$. 
We choose a six-site unit cell with positions specified as 
${\bf a}_1=(0,0)$, ${\bf a}_2=(0,1)$, ${\bf a}_3=(\sqrt3/2,3/2)$, 
${\bf a}_4=(\sqrt3,1)$, ${\bf a}_5=(\sqrt3,0)$, and ${\bf a}_6=(\sqrt3/2,-1/2)$. 
The supercell lattice vectors are given by ${\bf A}_{1/2}=(3\sqrt3/2,\pm3/2)$.

Within CDMFT as well as DCA, the interacting lattice Green's function in 
the site basis is defined as
\begin{equation}
     G_{ij}(i\omega_n) = \sum_{\bf k} \left[ i\omega_n + \mu - h({\bf k})- 
                   \Sigma(i\omega_n)\right]^{-1}_{ij} ,
\label{G}
\end{equation}
where $\omega_n=(2n+1)\pi T$ are Matsubara frequencies and $T$ is the temperature. 
At half-filling, the chemical potential is $\mu =U/2$. The ${\bf k}$ sum extends 
over the reduced Brillouin Zone, $h({\bf k})=-t({\bf k})$, where $t({\bf k})$ denotes 
the hopping matrix for the superlattice, and $\Sigma_{ij}(i\omega_n)$ represents 
the self-energy matrix in the site representation. 

Within CDMFT, the elements of $t({\bf k})$ within the unit cell given by 
$t_{ij}=t$ for neighboring sites. In addition, hopping between cells yields:  
\begin{eqnarray}
  t_{14} &=& t\, e^{-i{\bf k}\cdot {\bf A}_1}  \nonumber\\   
  t_{25} &=& t\, e^{-i{\bf k}\cdot {\bf A}_2}   \\   
  t_{36} &=& t\, e^{-i{\bf k}\cdot {\bf A}_3} ,  \nonumber
\end{eqnarray}   
where ${\bf A}_3={\bf A}_2-{\bf A}_1$. The hopping matrix $t({\bf k})$ is Hermitian, 
so that $t_{ji} = t^*_{ij}$. All other elements vanish.

To distinguish the hopping matrix elements within DCA, we denote them by
$\bar t_{ij}({\bf k})$. In the real-space version of DCA\cite{biroli} they are 
related to those within CDMFT via a phase factor:
\begin{equation}
    \bar t_{ij} = t_{ij}\, e^{-i{\bf k} \cdot {\bf a}_{ij}} ,
\label{phase}
\end{equation}
where ${\bf a}_{ij}={\bf a}_{i}-{\bf a}_{j}$. This phase relation yields the
following matrix elements:
\begin{eqnarray}
\bar t_{12}&=&\bar t_{36}=\bar t_{54}=t\,e^{-i{\bf k}\cdot{\bf a}_{12}}\nonumber\\   
\bar t_{23}&=&\bar t_{41}=\bar t_{65}=t\,e^{-i{\bf k}\cdot{\bf a}_{23}}  \\   
\bar t_{34}&=&\bar t_{52}=\bar t_{16}=t\,e^{-i{\bf k}\cdot{\bf a}_{34}},\nonumber
\end{eqnarray}   
with analogous connections among the Hermitian elements. All other matrix elements 
vanish.

\begin{figure}  [t!] 
\begin{center}
\includegraphics[width=5.0cm,height=8.5cm,angle=-90]{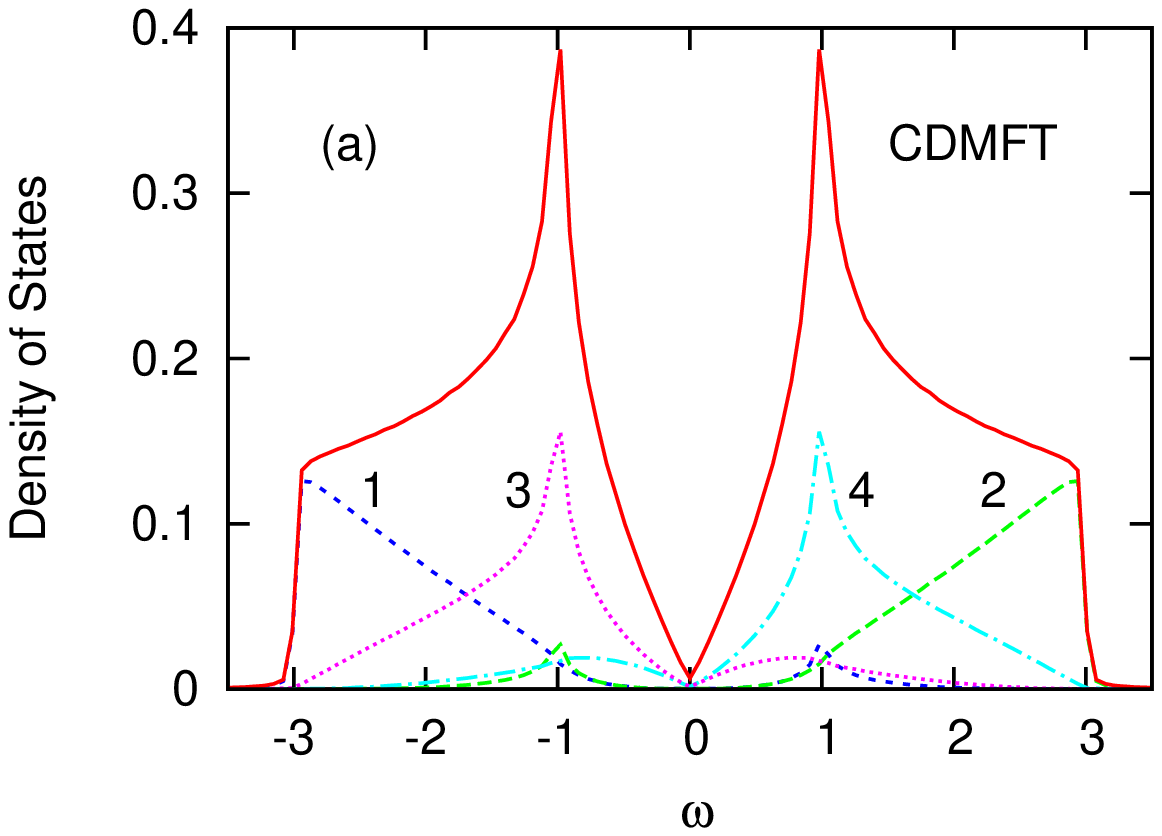}
\includegraphics[width=5.0cm,height=8.5cm,angle=-90]{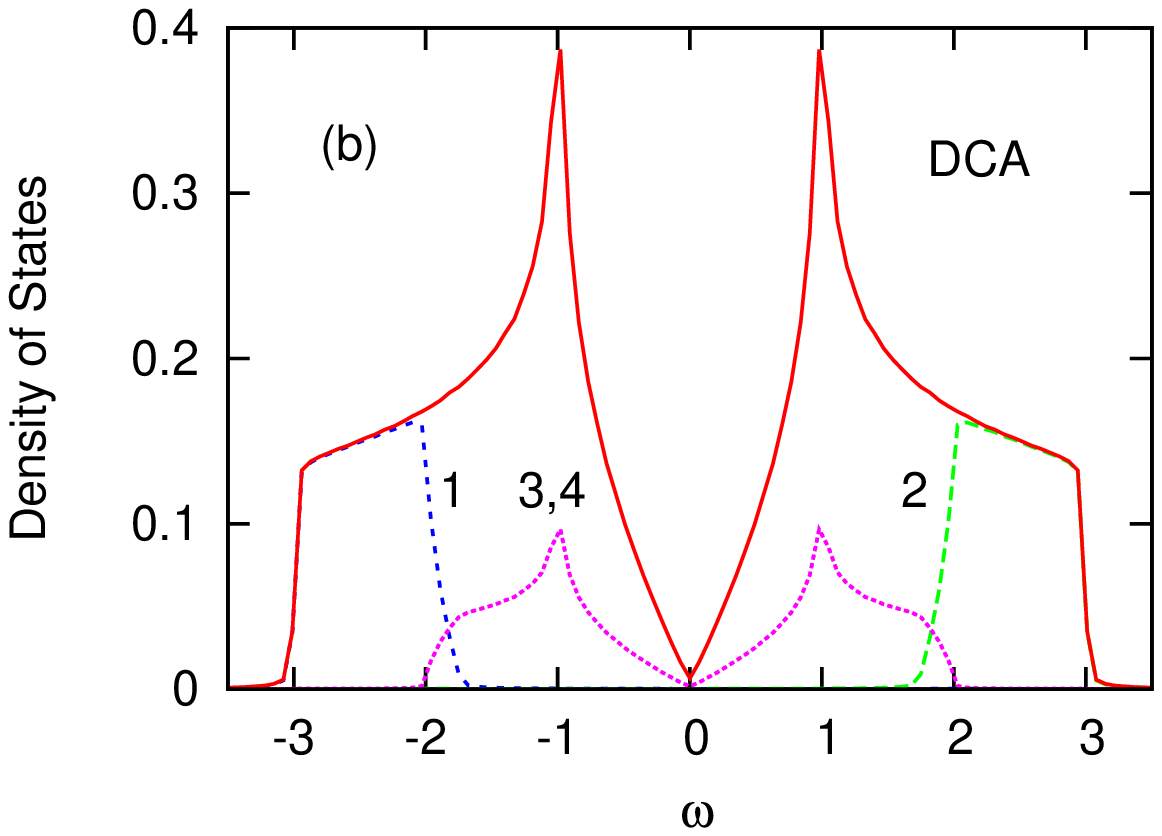}
\end{center}\vskip-3mm
\caption{(Color online)
Density of states $\rho(\omega)$ (solid curves) of honeycomb lattice and cluster 
components $\rho_m(\omega)$ in diagonal molecular orbital basis (dashed curves) 
for (a) CDMFT and (b) DCA. For clarity, these components are divided by $n_c=6$. 
In CDMFT, all density components are non-symmetric and orbitals 3 and 4 are doubly 
degenerate. In DCA, only $\rho_{1}$ and $\rho_{2}$ are non-symmetric, while the 
degenerate components $\rho_{m=3\ldots6}$ are symmetric. 
$\omega=0$ defines the Fermi energy for half-filling. 
}\label{dos}\end{figure}

\begin{figure}  [t!] 
\begin{center}
\includegraphics[width=5.0cm,height=6.0cm,angle=-90]{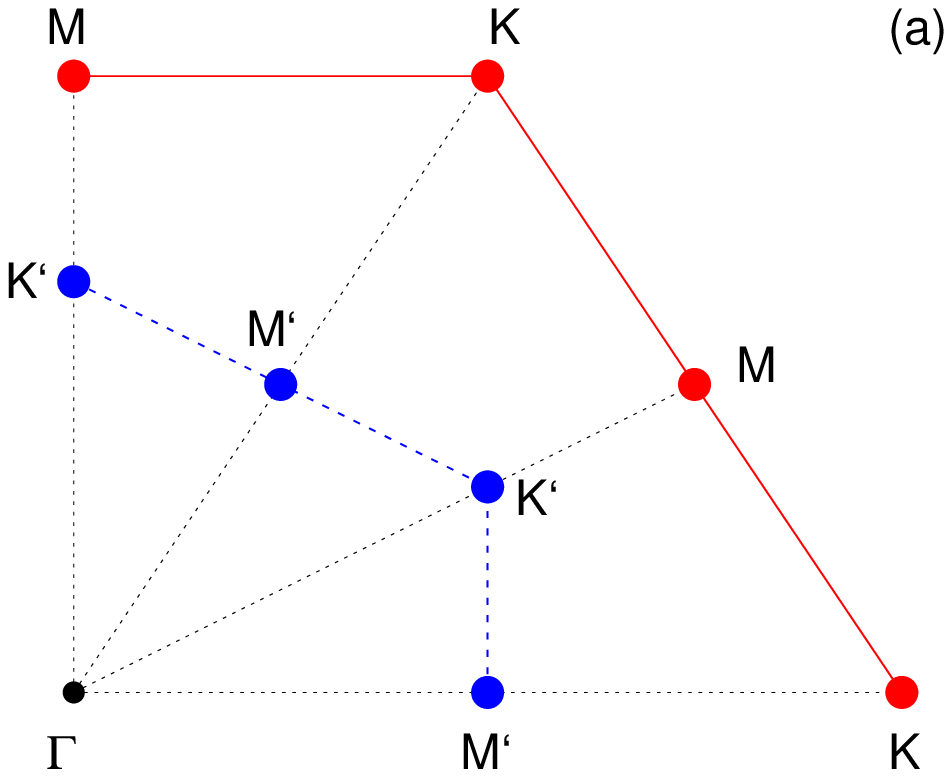}
\includegraphics[width=5.0cm,height=8.5cm,angle=-90]{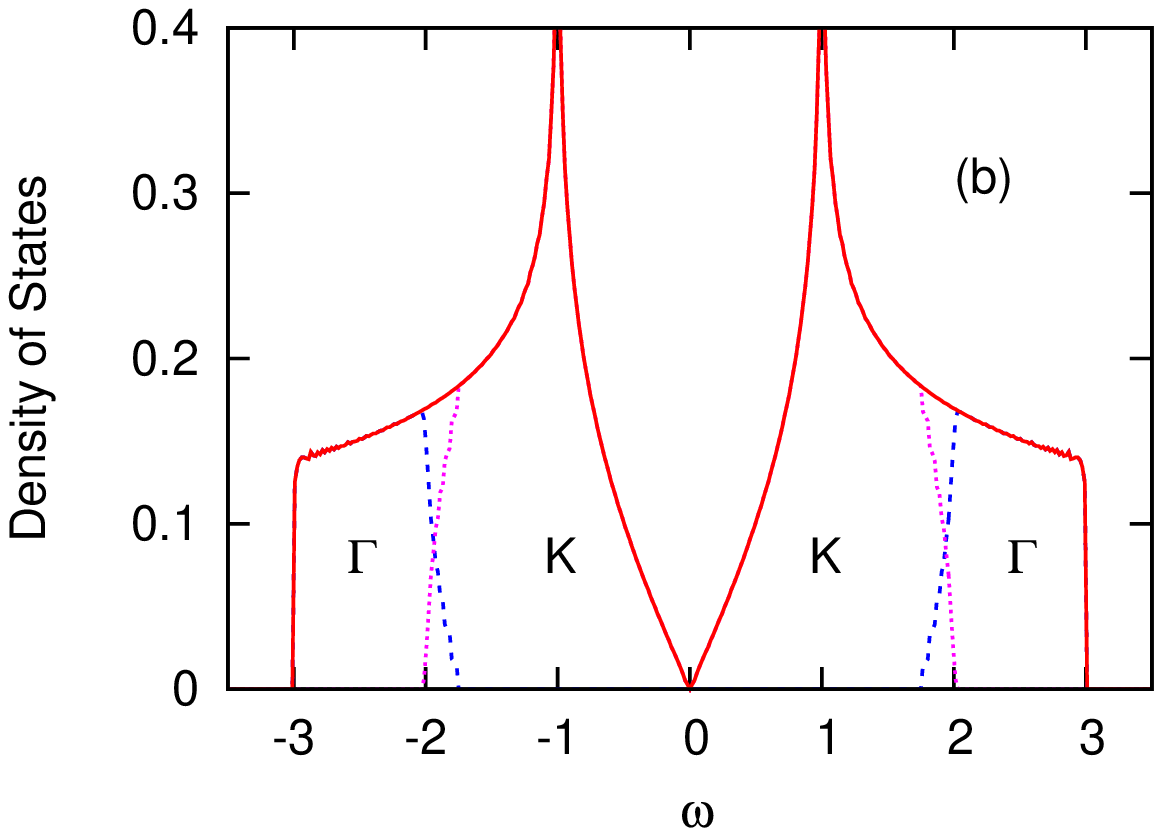}
\end{center}\vskip-3mm
\caption{(Color online)
(a) Segment of Brillouin Zone of honeycomb lattice (solid red lines). The reduced 
Zone (dashed blue lines) is obtained by folding the Dirac points $K$ onto $\Gamma$
and the $M$ points onto $M'$.      
(b) Decomposition of density of states into low-energy contribution (denoted as $K$)
corresponding to outer regions $KMK'M'$ and high-energy contribution (denoted as 
$\Gamma$) corresponding to inner regions $\Gamma M'K'$ of large Brillouin Zone.\\   
}\label{BZ}\end{figure}

The cluster Hamiltonian in CDMFT has the hopping matrix elements 
$[\sum_{\bf k} t({\bf k})]_{ij} = t^{cl}_{ij}$ where $t^{cl}_{ij}=t=1$ for 
first neighbors and $t^{cl}_{ij}=0$ otherwise. In contrast, in DCA we find
${\bar t}^{cl}_{ij}=\bar t=0.8103$ for first and third neighbors and 
${\bar t}^{cl}_{ij}=0$ otherwise. 

Within CDMFT as well as DCA, $G_{ij}$ is a symmetric matrix, with site-independent 
diagonal components $G_{ii}$. In the case of CDMFT, there are three independent 
off-diagonal elements: $G_{12}$, $G_{13}$ and $G_{14}$.  Here, $G_{11}$, $G_{13}$ 
are imaginary and $G_{12}$, $G_{14}$ are real. Thus, the corresponding density of 
states components $\rho_{11}$ and $\rho_{13}$ are even functions of energy, while 
$\rho_{12}$ and $\rho_{14}$ are odd. In the case of DCA, translation symmetry is 
preserved, so that one has the additional condition $\rho_{12}=\rho_{14}$ and 
$G_{12}=G_{14}$ due to the equality of first- and third-neighbor hopping 
interactions $\bar t$. 

Because of these symmetry properties, it is useful to express the lattice Green's 
function in the diagonal molecular-orbital basis whose elements $G_m(i\omega_n)$ 
($m=1\ldots6$) are determined by:
\begin{eqnarray}
 G_{1,2} &=& (G_{11}+2G_{13}) \pm (G_{14}+2G_{12}) \nonumber \\ 
 G_{3,4} = G_{5,6}                               
         &=& (G_{11}- G_{13}) \pm (G_{14}-G_{12}).  \label{Gk}
\end{eqnarray}
The unitary transformation $\bar T_{im}$ linking the site and molecular-orbital 
bases is defined in Eq.~(6) of Ref.~\onlinecite{PRB2011}.
Evidently, in CDMFT there are two independent complex functions, $G_1=-G^*_2$ and 
$G_3=-G^*_4$. In DCA, the elements $G_{m=3\ldots6}$ are degenerate and imaginary.
The onsite and intersite components of the lattice Green's function can be derived 
by inverting Eq.~(\ref{Gk}):
\begin{eqnarray}
 G_{11} &=& [(G_1 + G_2)  + 2 (G_3 + G_4)]/6 \nonumber \\ 
 G_{12} &=& [(G_1 - G_2)  -   (G_3 - G_4)]/6   \nonumber\\
 G_{13} &=& [(G_1 + G_2)  -   (G_3 + G_4)]/6   \nonumber \\ 
 G_{14} &=& [(G_1 - G_2)  + 2 (G_3 - G_4)]/6 .  
\end{eqnarray}

Figure~\ref{dos} illustrates the uncorrelated density of states components in the 
diagonal molecular orbital basis, where 
$\rho_m(\omega) = -\frac{1}{\pi}\,{\rm Im}\, G_m(\omega)$.
The total density of states is, of course, the same within CDMFT and DCA, but its
decomposition into molecular-orbital or intersite contributions differs for these
two schemes.
The four CDMFT densities shown in panel (a) are non-symmetric and satisfy the 
relations $\rho_2(\omega)=\rho_1(-\omega)$ and $\rho_4(\omega)=\rho_3(-\omega)$.
The corresponding DCA densities are plotted in panel (b). In this case, only 
$\rho_1(\omega)=\rho_2(-\omega)$ are non-symmetric, whereas 
$\rho_{3}(\omega)=\rho_{4}(\omega)$ are symmetric. 

Figure~\ref{BZ} (a) shows the Brillouin Zone of the honeycomb lattice together  
with the three times smaller reduced Zone. Panel (b) illustrates the contributions 
to the density of states stemming from the outer ${\bf k}$ regions $KMK'M'$ and 
the inner regions $\Gamma M'K'$. These two contributions overlap slightly since 
the point $K'$ does not lie half-way between $\Gamma$ and $M$. Thus,
the low-energy part of the density of states (denoted as $K$) extends up 
$\vert \omega\vert \le 2$, while the high-energy part (denoted as $\Gamma$) 
corresponds to the window $1.75\le\vert\omega\vert\le3$. The comparison with 
Fig.~\ref{dos} (b) shows that the diagonal elements of the DCA density of states 
correspond to the distributions indicated in Fig.~\ref{BZ} (b). Thus, 
$\rho_{1,2}(\omega)$ account for the energy bands in the inner regions $\Gamma M'K'$ 
and  $\rho_{3,4}(\omega)$ for those in the outer regions $KM K'M'$ of the original 
Brillouin Zone. The momentum regions shown in Fig.~\ref{BZ} (a) therefore specify 
the appropriate tiling of the Brillouin Zone within the DCA.   
        
The self-energy matrices in CDMFT and DCA satisfy the same symmetry properties as 
the lattice Green's functions so that they can be diagonalized in the same manner. 
These diagonal elements will be denoted as $\Sigma_m(i\omega_n)$. In the site basis 
the components $\Sigma_{11}$ and $\Sigma_{13}$ are imaginary, whereas $\Sigma_{12}$ 
and $\Sigma_{14}$ are real. As translation symmetry is not obeyed in CDMFT, 
$\Sigma_{12}$ and $\Sigma_{14}$ differ, while in DCA they coincide. 

We point out that, although the hopping matrix elements $t({\bf k})$ in CDMFT and 
DCA differ only by a unitary transformation as indicated in Eq.~(\ref{phase}), the
same does not hold for the respective self-energy matrices. As discussed below, 
the preservation of translation invariance in DCA and its absence in CDMFT give 
rise to fundamentally different physical solutions which severely affect the phase 
boundaries. Thus, the DCA and CDMFT self-energy matrices are not simply related
via a unitary transformation.     

Severe differences of this kind between DCA and CDMFT do not arise in the case of 
the Hubbard model for the square lattice, where the cluster Hamiltonians maintain 
the same symmetry. The only difference is that the hopping interaction between 
neighbors is changed from $t=1$ in CDMFT to $\bar t=1.273$ in DCA. As a result, 
these cluster schemes lead to a less dramatic reorganization of spectral weight 
among the cluster components than in the case of the honeycomb lattice.

\section{Exact Diagonalization}

\begin{figure}  [t!] 
\begin{center}
\includegraphics[width=5.5cm,height=6.0cm,angle=-90]{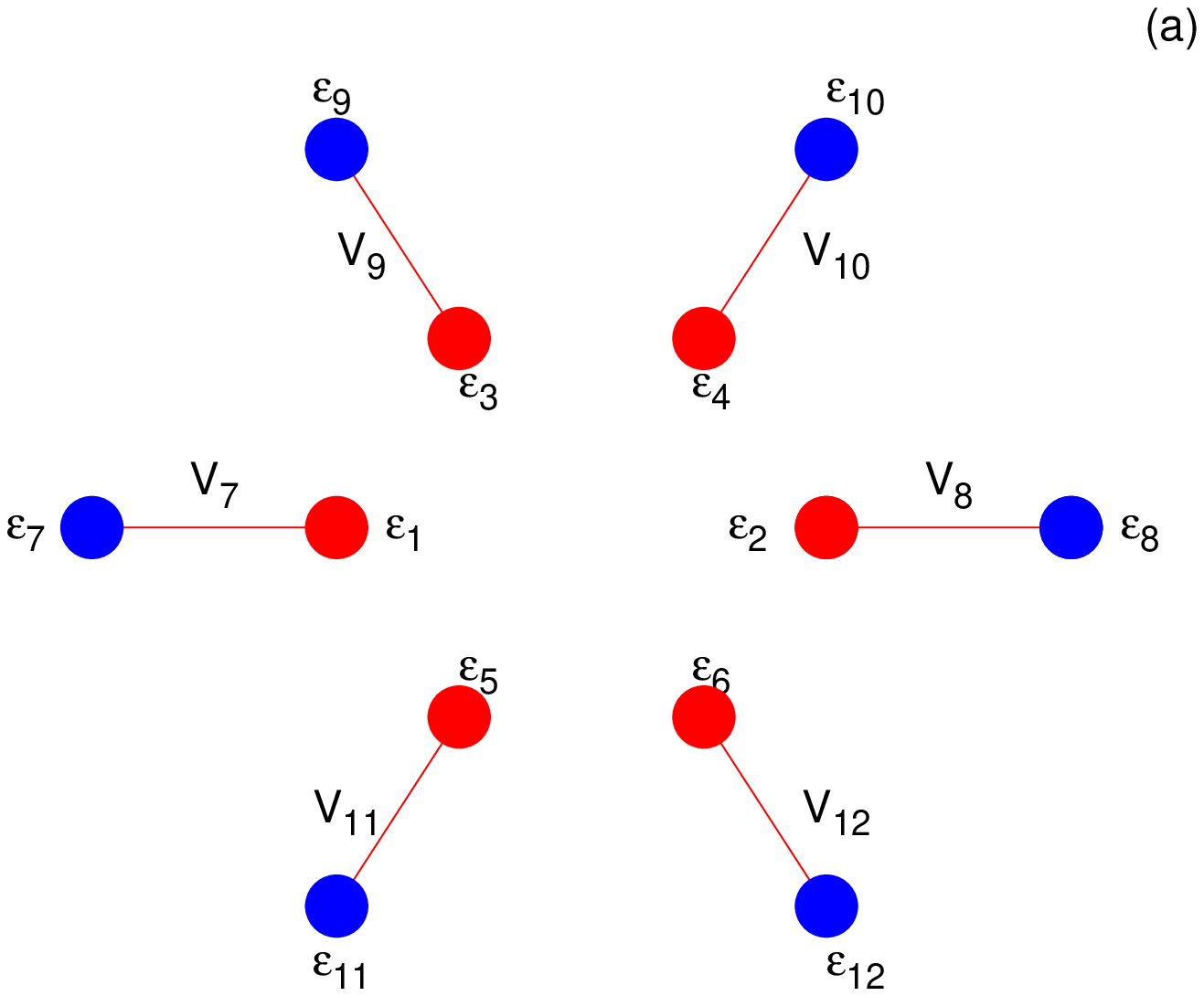}\vskip-5mm
\includegraphics[width=5.5cm,height=6.0cm,angle=-90]{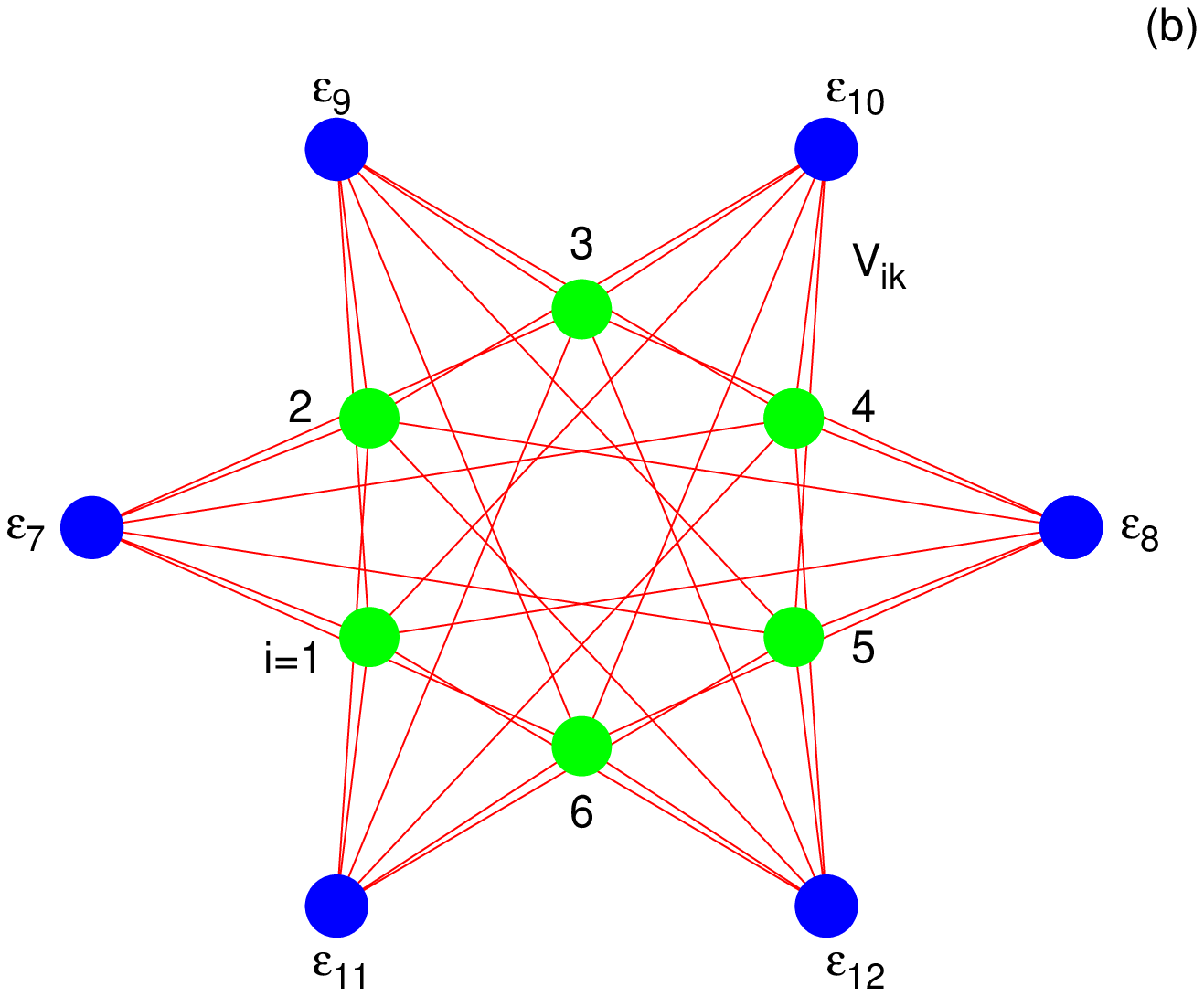}
\end{center}\vskip-3mm
\caption{(Color online)
(a) Cluster levels in molecular orbital basis. There are six independent terms
connecting orbital levels $\epsilon_{m=1...6}$ (red dots) and bath levels 
$\epsilon_{k=7...12}$ (blue dots) via hopping integrals $V_{k=7...12}$. In CDMFT 
(for fixed impurity levels) one has: 
$\epsilon_{1,2}=\mp 2t$, $\epsilon_{3,4}=\epsilon_{5,6}=\pm t$, and 
$\epsilon_7=-\epsilon_8$, $\epsilon_9=\epsilon_{11}=-\epsilon_{10}=-\epsilon_{12}$, 
$V_7=V_8$, $V_9=V_{10}=V_{11}=V_{12}$.
Thus there are four independent bath parameters. 
In DCA, $ \epsilon_{1,2}=\mp 3\bar t$, $\epsilon_{3...6}=\epsilon_{9...12}=0$, 
{\it i.e.}, there are only three independent fit parameters.
(b) Cluster levels in site basis $i=1...6$ (green dots) connected to molecular 
orbital bath levels $\epsilon_{k=7...12}$ (blue dots) via hopping integrals $V_{ik}$. 
For clarity, the hopping interactions between impurity sites are not shown. 
Representations (a) and (b) are equivalent since they are connected via the unitary
transformation $\bar T$ between impurity sites $i=1...6$ and orbitals $m=1...6$. 
The bath molecular orbital levels in (b) are the same as in (a). Thus, although the
cluster sites have identical levels at zero energy, the bath levels maintain the 
orbital symmetry.      
}\label{levels}\end{figure}

To avoid double-counting of Coulomb interactions in the quantum impurity calculation, 
the self-energy must be removed from the six-site cluster in which correlations are 
treated explicitly. This removal yields the bath Green's function matrix 
\begin{equation}
         G_0(i\omega_n) = [G(i\omega_n)^{-1} + \Sigma(i\omega_n)]^{-1} .  
           \label{G0}
\end{equation} 
Within the ED approach, this bath Green's function of the infinite lattice
is projected onto the corresponding function of a supercluster consisting of 
$n_c=6$ correlated sites within the unit cell plus a bath consisting of $n_b$ 
discrete levels. Here, we choose $n_b=6$, so that the total number
of levels of the supercluster is $n_s=n_c+n_b=12$. Within the diagonal 
molecular-orbital basis, this projection implies 
\begin{eqnarray}
 G_{0,m}(i\omega_n) &\approx&  G^{cl}_{0,m}(i\omega_n) \nonumber\\
    &=&    \left( i\omega_n + \mu -\epsilon_m -
 \sum_{k=7}^{12} \frac{\vert V_{mk}\vert^2}{i\omega_n - \epsilon_k}\right)^{-1},
   \label{G0m}
\end{eqnarray}
where $\epsilon_{m=1...6}$ denotes impurity levels and $\epsilon_{k=7...12}$ bath 
levels. The bath levels are defined relative to the chemical potential. We assume
that the molecular orbitals couple to independent baths so that the hybridization 
matrix elements are also diagonal in this representation: $V_{mk}=\delta_{m+6,k}V_k$.
Fig.~\ref{levels}(a) illustrates the impurity and bath levels in the diagonal
molecular orbital basis. Panel (b) shows the equivalent representation when the
impurity orbitals are transformed to the original site basis. The bath remains 
unchanged and the hopping terms in this basis are given by 
$V_{ik} =\sum_m \bar T_{im} V_{mk}$.  
This picture differs from the one in which also the bath is treated 
within the site basis (see below).

To determine the bath levels $\epsilon_k$ and hopping terms $V_{mk}$
we minimize the difference
\begin{equation}
 {\rm Diff}_m = \sum_{n=0}^M  W_n^N
       \vert G_{0,m}(i\omega_n) - G^{cl}_{0,m}(i\omega_n)\vert^2  ,
\label{diff}
\end{equation}
where $M\approx 2^{10}$ is the total number of Matsubara points and the weight
function  $W^N_n=1/\omega^N_n$ is introduced to give more weight to the low-frequency
region. We usually take $N=1$ or $N=2$. Note also that both Green's functions in the 
above expression approach $1/i\omega_n$ for large $\omega_n$. 
Thus the difference defined in Eq.~(\ref{diff}) automatically focuses on the low-energy 
region. This is not the case when the differences of the inverse Green's functions are
minimized instead. The reason is that the hybridization functions corresponding to
$G_{0,m}$ and $G_{0,m}^{cl}$ are not normalized to the same asymptotic amplitudes. 
To start the iterative procedure, we use bath parameters obtained for the uncorrelated 
system, or from a converged solution for nearby Coulomb energies. The resulting 
$\epsilon_k$ and $V_{mk}$ are usually very stable against variations of initial 
conditions.

In the CDMFT calculations discussed in Ref.~\onlinecite{PRB2011}, not only the bath 
levels $\epsilon_k$ and hopping elements $V_k$ were used as parameters in the fit of
$G_{0,m}(i\omega_n)$, but also the impurity levels $\epsilon_m$. Since the 
expression Eq.~(\ref{G0m}) ensures the correct asymptotic behavior, the variation
of $\epsilon_m$ yields slightly better accuracy of the fit at the lower Matsubara
points. For each diagonal component $G_{0,m}$ three fit parameters are then available. 
As there are only two independent complex functions $G_{0,m}$, the total number of 
parameters to fit the bath is six. As shown in Fig.~24 of Ref.~\onlinecite{EDR} 
for $U=4$ and $T=0.01$, this procedure yields a surprisingly good reproduction 
of the lattice bath Green's function via the cluster Anderson Green's function,
in spite of the fact that we use only one bath level per impurity orbital. 
The reason for this good fit is the semi-metallic nature of the honeycomb lattice,
giving rise to a vanishing density of states at the Fermi level. In contrast, in 
ordinary correlated metals and the triangular or square lattice Hubbard models, 
the density of states of the infinite lattice is finite, so that a successful fit 
to a cluster Green's function usually requires at least two bath levels per orbital 
and restriction to not very low temperatures (typically $T\ge0.01$). 

In the DCA calculations presented below, we fix the impurity levels $\epsilon_m$ 
at their nominal cluster values. Thus, $\epsilon_{1,2}=\mp 3\bar t$ and 
$\epsilon_{3,4}=0$. The latter value reflects the fact that the DCA density of 
states components $\rho_{3,4}(\omega)$ are even functions of energy. Thus, the fit 
of $G_{0,m=1,2}$ involves two parameters (the bath level $\epsilon_7=-\epsilon_8$ 
and the hopping element $V_7=V_8$), while $G_{0,m=3,4}$ includes only the hopping 
element $V_9=V_{10}$ as fit parameter.

\begin{figure}  [t!] 
\begin{center}
\includegraphics[width=6.5cm,height=8.5cm,angle=-90]{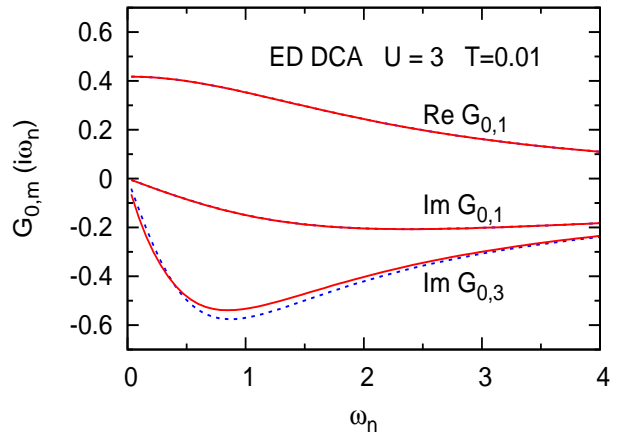}
\end{center}\vskip-3mm
\caption{(Color online)
Comparison of lattice bath Green's function $G_{0,m}(i\omega_n)$ (solid red curves)
and cluster Green's function (dashed blue curves) for $U=3$ and $T=0.01$. 
As the density of states for $m=3$ is symmetric in DCA (see Fig.~1 (b)), $G_{0,3}$ 
is purely imaginary, while $G_{0,1}$ is complex. Thus, the latter function is 
fitted with two parameters, whereas $G_{0,3}$ involves only one fit parameter.  
The solid and dashed curves for $G_{0,1}$ are indistinguishable.       
}\label{G0.DCA}\end{figure}

Figure \ref{G0.DCA} illustrates the quality of the fit of $G_0$ within ED DCA for 
$U=3$ and $T=0.01$. The parameters used in these fits are: 
$\epsilon_1=-3\bar t=-2.4309$,
$\epsilon_7=-1.85694$, $V_7=0.26270$ for $m=1$ and $\epsilon_3=\epsilon_9=0$,
$V_9=0.86701$ for $m=3$.         
As pointed out above in the case of CDMFT, the excellent representation of the 
lattice Green's function via the cluster Green's function using only one bath level 
per impurity orbital is related to the vanishing density of states at the Fermi level. 
  
The diagonalization of the supercluster Hamiltonian is conveniently carried out in 
the site basis. At low temperatures only few excited states need to be included 
in the evaluation of the cluster Green's function $G^{cl}_{ij}(i\omega_n)$. The 
diagonalization can then be performed very efficiently by making use of the Arnoldi 
algorithm. Details concerning this procedure are provided in 
Refs.~\onlinecite{perroni,tong,EDR}.
Since the cluster Green's function obeys the same symmetry properties as the 
lattice Green's function, it is diagonal in the molecular-orbital basis. These
elements will be denoted as $G^{cl}_m(i\omega_n)$. The diagonal cluster self-energy 
components are then given by an expression analogous to Eq.~(\ref{G0}):
\begin{equation}
\Sigma^{cl}_{m}(i\omega_n) = 1/G^{cl}_{0,m}(i\omega_n)-1/G^{cl}_{m}(i\omega_n) .
\label{Scl}
\end{equation} 
The key physical assumption in DMFT is now that this cluster self-energy provides 
an accurate representation of the lattice self-energy. Thus, 
\begin{equation}
     \Sigma_{m}(i\omega_n) \approx \Sigma^{cl}_{m}(i\omega_n) .
\label{S}
\end{equation}
In the next iteration, these self-energy components are used as input in the 
lattice Green's function Eq.~(\ref{G}). In the diagonal molecular-orbital basis 
the DCA lattice Green's function is given by 
\begin{equation}
G_{m}(i\omega_n)=\sum_{\bf k}\left[i\omega_n+\mu-{\bar T}^{-1}\bar h({\bf k}){\bar T} 
                    -  \Sigma(i\omega_n)\right]^{-1}_{mm} .
\label{Gm}
\end{equation}
We note here that, at real energies, the cluster quantities $G^{cl}_{m}$, 
$G^{cl}_{0,m}$ and $\Sigma^{cl}_{m}$ have discrete spectra, while the corresponding
lattice spectra associated with the quantities $G_{m}$, $G_{0,m}$ and $\Sigma_{m}$ 
are continuous. 

We close this section by pointing out that we believe the projection of the bath 
Green's function within the diagonal molecular-orbital basis discussed above to be 
more general and more flexible than analogous projections within the nondiagonal site 
basis. As mentioned above, within CDMFT there are two independent complex functions 
$G_{0,m}$ (with nonsymmetric spectral distributions) that are fitted each with
one bath level $\epsilon_k$ and one hopping term $V_k$ (assuming the impurity 
level $\epsilon_m$ to be fixed). Thus, there are altogether four fit parameters. 
This should be compared to only one fit parameter if the site basis is used instead.
For symmetry reasons all bath levels then are zero so that only the site independent
impurity bath hopping element remains as a single fit parameter. Introducing a 
hopping interaction among bath levels as was done in Ref.~\onlinecite{helu} 
increases the number of fit parameters from one to two. 
Actually, since the bath can always be represented in a diagonal form, hopping 
among bath levels is implicitly included in the diagonal molecular orbital picture 
with four fit parameters. Analogous considerations hold for DCA.        
Nevertheless, as will be shown in the next section, these slightly different 
implementations of ED all yield consistent answers concerning the variation of
the excitation gap as a function of Coulomb energy.

\section{Results and Discussion}

\begin{figure}  [t!] 
\begin{center}
\includegraphics[width=6.0cm,height=8.5cm,angle=-90]{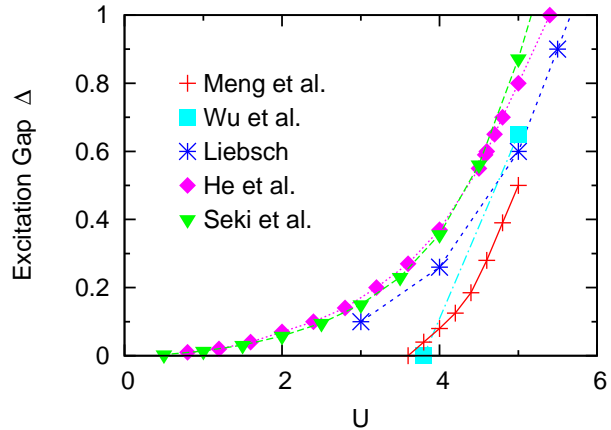}
\end{center}\vskip-3mm
\caption{(Color online)
Comparison of excitation gaps as functions of Coulomb interaction derived using 
several cluster methods and impurity solvers for paramagnetic phase of honeycomb 
lattice:  
Meng {\it et al.}: large-scale QMC\cite{meng}, 
Wu   {\it et al.}: CTQMC CDMFT    \cite{wu}, 
Liebsch:              ED CDMFT    \cite{PRB2011},
He   {\it et al.}:    ED CDMFT    \cite{helu}, 
Seki {\it et al.}:    ED VCA      \cite{seki}. 
In contrast, both ED and CTQMC DCA yield semi-metallic behavior with $\Delta=0$
for $U\le 6$ (see text).
}\label{gap}\end{figure}

Figure~\ref{gap} shows the comparison of the excitation gaps obtained for various
cluster methods and impurity solvers. Near $U\approx5$, all calculations (except 
DCA, see below) predict a Mott 
phase with a gap $\Delta\approx 0.5-0.9$. At $U\le4$, the CDMFT and VCA results
that do not preserve translation symmetry exhibit a gap tail that persists
down to $U\rightarrow0$. The differences between these results are partly caused 
by the different temperatures used in these studies. In particular, the gap 
closing near $U=3.8$ obtained within CDMFT by Wu {\it et al.}\cite{wu} seems 
to be related to the rather high temperature, $T=0.05$, employed in the CTQMC
calculation. Since the CTQMC self-energy agrees well with the ED results, 
CTQMC CDMFT presumably would also yield a gap at lower $T$. 
Also, the ED calculations in Ref.~\onlinecite{PRB2011} were carried out at 
$T=0.005$, while those in Refs.~\onlinecite{helu,seki} essentially correspond
to the $T\rightarrow0$ limit.         

In striking contrast to CDMFT, the translation invariance of DCA ensures the 
existence of a semi-metallic phase at low values of $U$. 
On the other hand, the condition $\Sigma_{12}=\Sigma_{14}$ cannot generally be 
correct for the short-range correlations within the unit cell. Thus, at Coulomb
energies, where local Mott physics dominates and long-range translational 
invariance becomes less important, DCA should be less appropriate than CDMFT.
Indeed, both ED and CTQMC DCA results suggest that the semi-metallic phase with 
$\Delta=0$ extends to $U>6$, i.e., beyond the critical Coulomb energy 
$U_c\approx 3.9-4.3$ of the antiferromagnetic phase.\cite{meng,sorella}  

\begin{figure}  [t!] 
\begin{center}
\includegraphics[width=6.5cm,height=8.5cm,angle=-90]{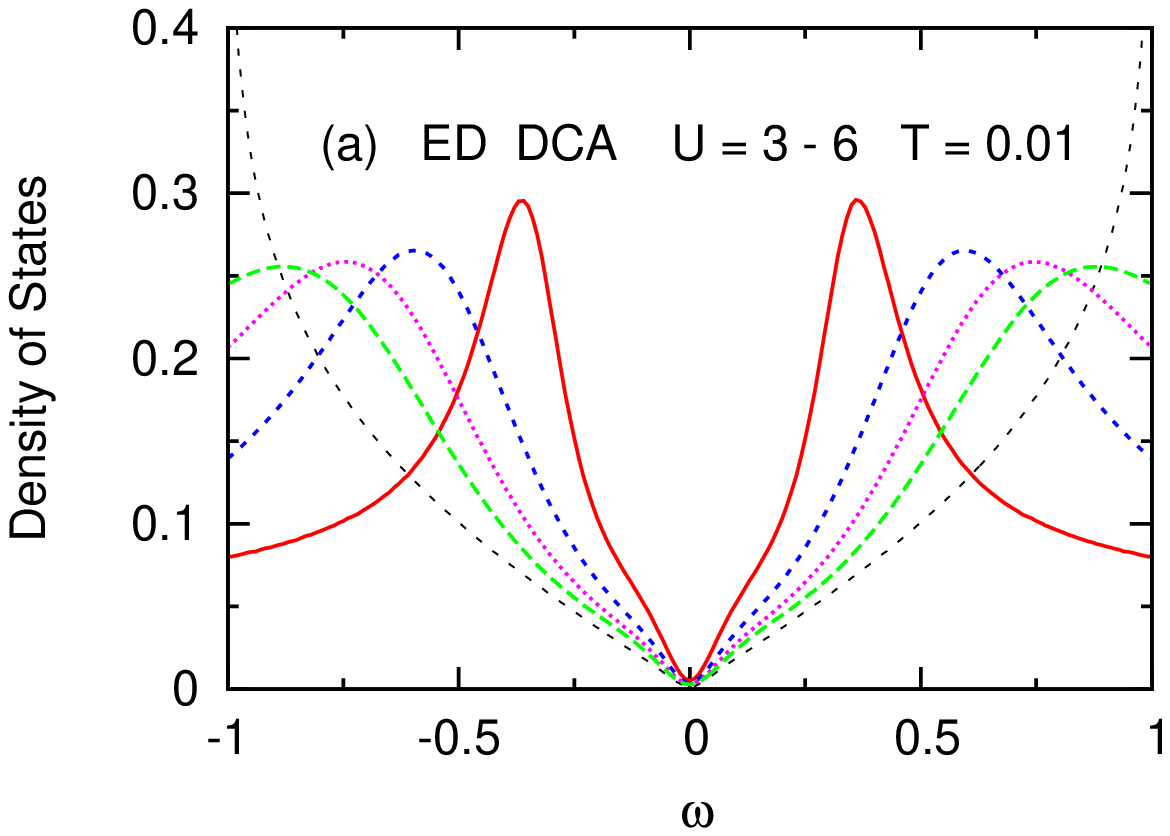}
\includegraphics[width=6.5cm,height=8.5cm,angle=-90]{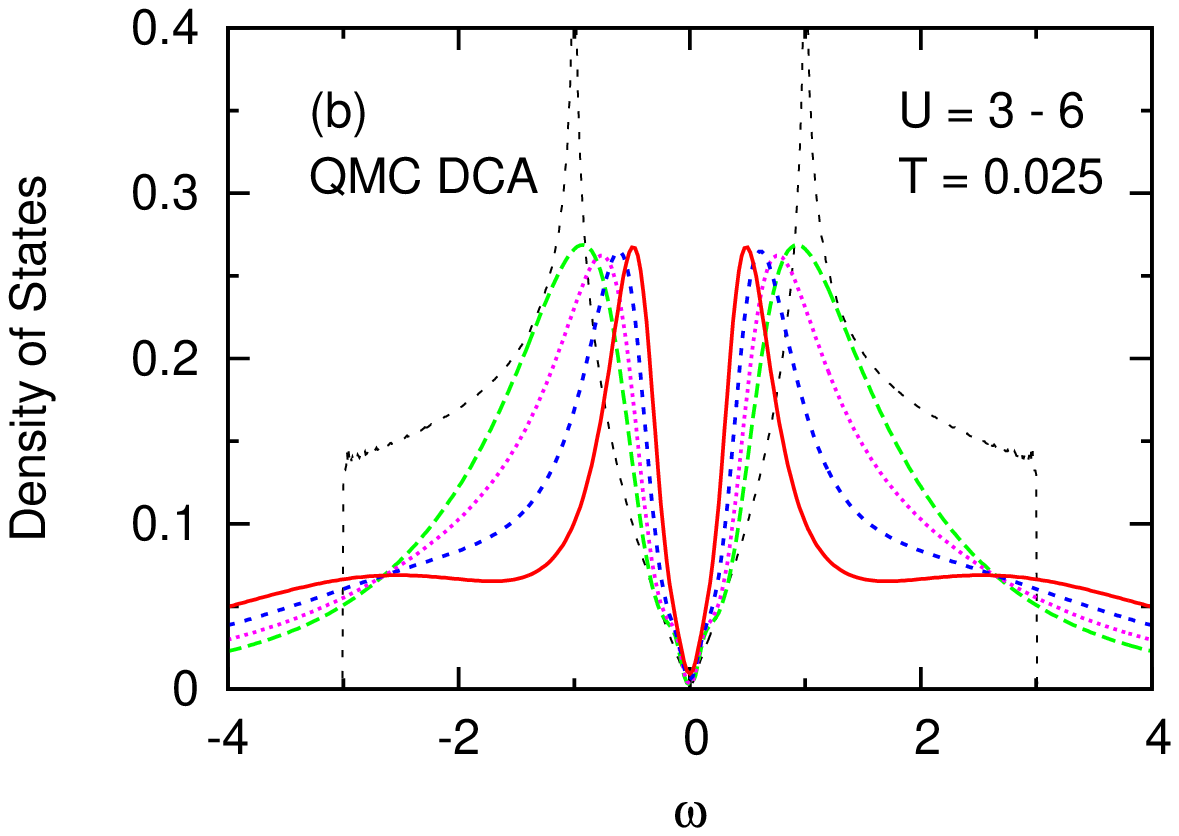}
\end{center}\vskip-3mm
\caption{(Color online)
Density of states $A_{11}(\omega)=-\frac{1}{\pi}{\rm Im}G_{11}(\omega)$ 
of honeycomb lattice for several Coulomb energies at. Red solid curves:
$U=6$, dashed curves: $U=3 - 5$. (a) ED DCA ($T=0.01$) (b) CTQMC DCA ($T=0.025)$.   
For illustrative purpose, only the low-energy range of the ED spectra is shown.
The dotted curve denotes the bare density of states.
}\label{spectra}\end{figure}

This is illustrated in Fig.~\ref{spectra}, which shows the interacting density 
of states obtained in ED and CTQMC DCA for several Coulomb energies. 
The ED spectra were obtained by making use of the extrapolation routine 
{\it ratint}, \cite{ratint} while the CTQMC spectra were derived via the 
maximum entropy method.\cite{maxent} 
For details concerning the CTQMC calculations, see Ref.~\onlinecite{wu}. 
The main effect of Coulomb interactions 
is seen to be the usual band narrowing and effective mass enhancement, as found 
in weakly correlated systems. In contrast, the corresponding ED and CTQMC 
CDMFT spectra for $U=5$ reveal a large Mott gap of about $\Delta=0.6$ 
(see Fig.~\ref{gap}).\cite{wu,PRB2011}  

The persistence of semi-metallic behavior at large $U$ within DCA is related 
to the fact that the enforcement of translation symmetry is achieved at the 
expense of equating first- and third-neighbor interactions in the cluster 
Hamiltonian. The self-energy in the site basis then satisfies the condition 
$\Sigma_{12}=\Sigma_{14}$, whereas in CDMFT $\Sigma_{14}$ is noticeably smaller 
than $\Sigma_{12}$.\cite{PRB2011,EDR}

\begin{figure}  [t!] 
\begin{center}
\includegraphics[width=6.5cm,height=8.5cm,angle=-90]{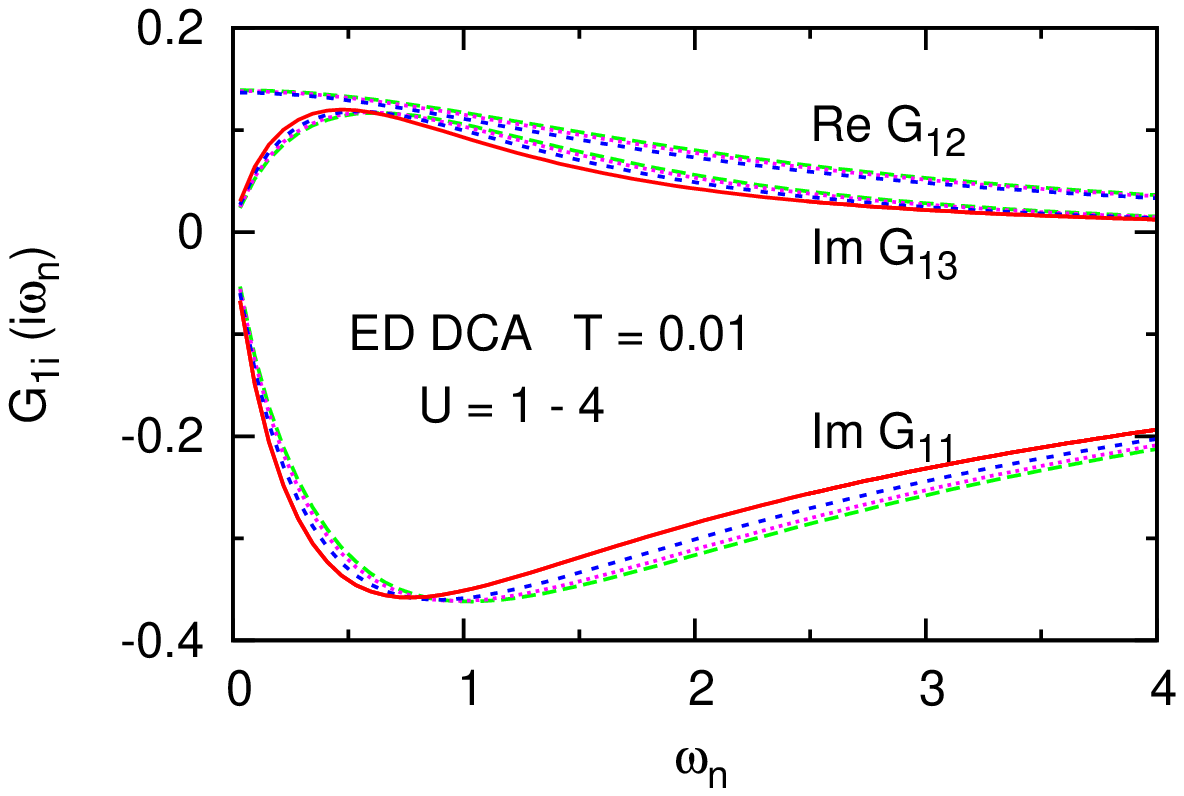}
\includegraphics[width=6.5cm,height=8.5cm,angle=-90]{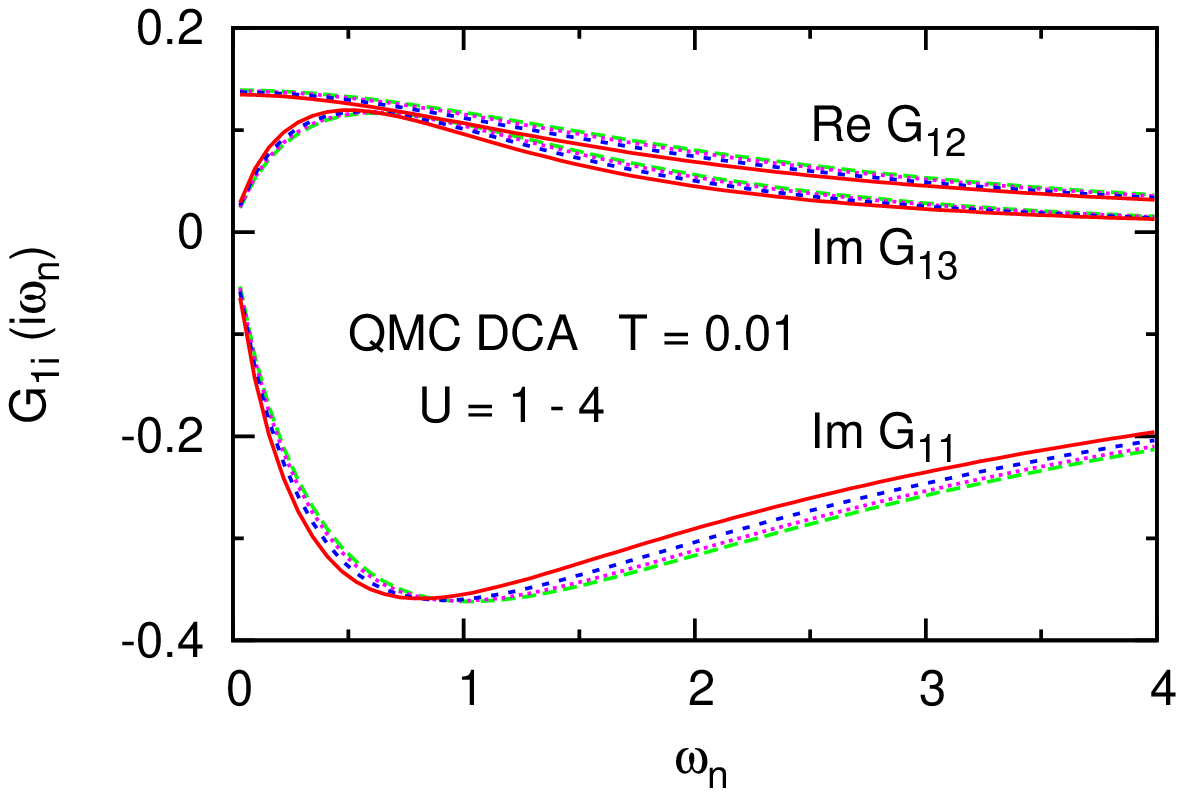}
\end{center}\vskip-3mm
\caption{(Color online)
Green's function components $G_{1i}(i\omega_n)$ ($i=1,2,3$) of honeycomb lattice 
as functions of Matsubara frequency calculated within (a) ED DCA and (b) CTQMC DCA
at $T=0.01$. Solid red curves: $U=4$; dashed curves: $U=1 - 3$.     
}\label{Gij}\end{figure}

\begin{figure}  [t!] 
\begin{center}
\includegraphics[width=6.5cm,height=8.5cm,angle=-90]{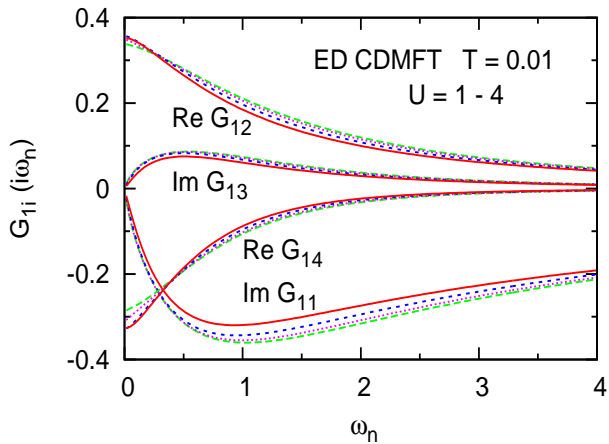}
\end{center}\vskip-3mm
\caption{(Color online)
Green's function components $G_{1i}(i\omega_n)$ ($i=1,2,4$) of honeycomb lattice 
as functions of Matsubara frequency calculated within ED CDMFT.
Solid red curves: $U=4$; dashed curves: $U=1 - 3$.     
}\label{Gij.CDMFT}\end{figure}

The good correspondence between the DCA spectra obtained within ED and CTQMC is a 
consequence of the nearly quantitative agreement of the lattice Green's functions 
$G_{1i}(i\omega_n)$ which are shown in Fig.~\ref{Gij}. As pointed out in the 
preceding section, for symmetry reasons $G_{11}$ and $G_{13}$ are imaginary, 
while $G_{12}=G_{14}$ are real. Both impurity solvers yield 
Im\,$G_{11}(i\omega_n)\rightarrow0$
in the limit $\omega_n\rightarrow 0$, implying that the local density of states,
$\rho(\omega)= -\frac{1}{\pi}\,{\rm Im}\,G_{11}(\omega)$ vanishes at $\omega=0$.  
Also, both schemes indicate that with increasing values of $U$ the initial
slope of Im\,$G_{11}$ and Im\,$G_{13}$ increases. Thus, the Dirac cones become 
steeper and spectral weight is shifted towards the Fermi level.        

The results obtained within DCA differ in two qualitative aspects from those
derived previously in CDMFT. As shown in Fig.~\ref{Gij.CDMFT}, the components 
$G_{12}$ and $G_{14}$ in CDMFT do not coincide.
Moreover, the initial slopes of Im\,$G_{11}$ and Im\,$G_{13}$ become smaller
with increasing $U$ rather than larger as within DCA. In Ref.~\onlinecite{PRB2011} 
it was demonstrated that for $U\ge 4$ a Mott gap opens in the density of states,
in approximate agreement with the large-scale QMC calculations by Meng {\it et al.}
\cite{meng} At smaller values of $U$, a tiny gap or pseudogap was also found (see
below), which is however difficult to resolve within ED at finite $T$. As the opening 
of a gap in the density of states implies a reduction of 
$\vert{\rm Im}\,G_{11}(i\omega_n)\vert$ 
at small values of $\omega_n$, the results shown in Figs.~\ref{Gij} and \ref{Gij.CDMFT}
underline the fundamental difference between DCA and CDMFT for the honeycomb lattice:
Whereas DCA yields a weakly correlated semi-metal, CDMFT gives rise to insulating
behavior even at small $U$.           

\begin{figure}  [t!] 
\begin{center}
\includegraphics[width=5.0cm,height=8.5cm,angle=-90]{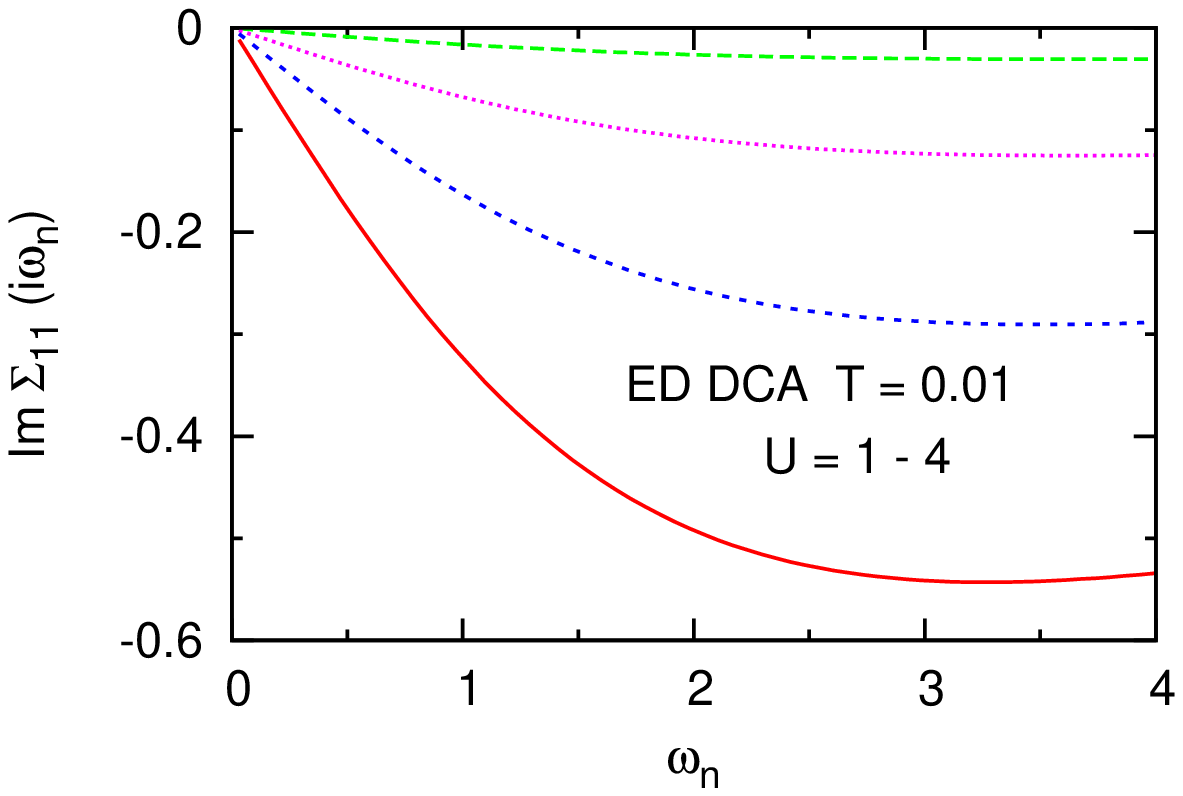}
\includegraphics[width=5.0cm,height=8.5cm,angle=-90]{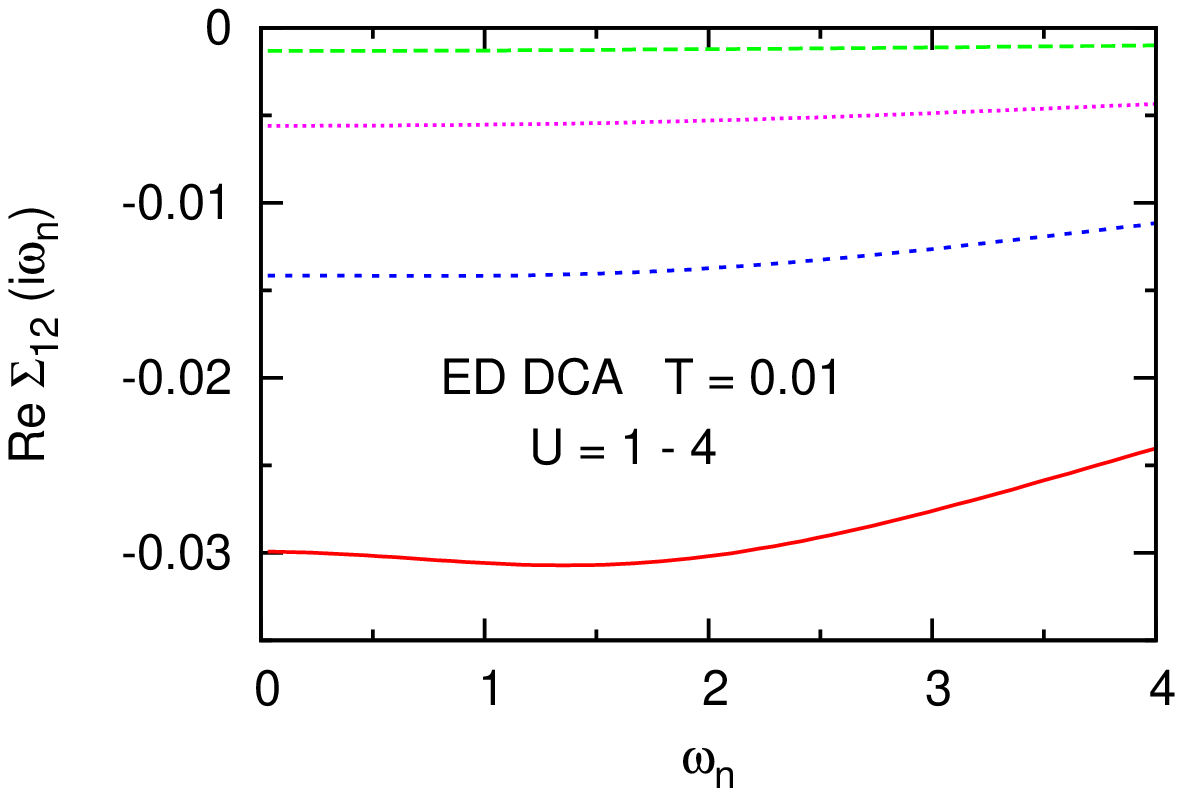}
\includegraphics[width=5.0cm,height=8.5cm,angle=-90]{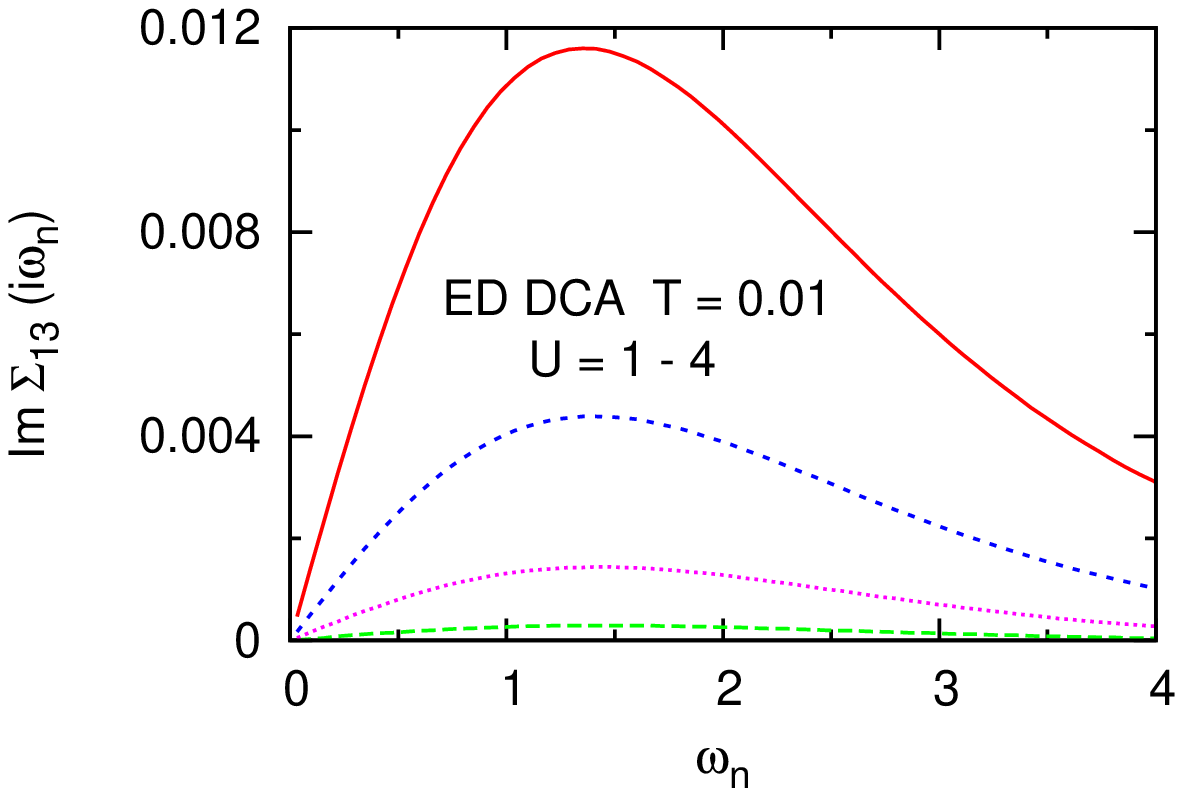}
\end{center}\vskip-3mm
\caption{(Color online)
Self-energy components $\Sigma_{1i}(i\omega_n)$ ($i=1,2,3$) of honeycomb lattice
as functions of Matsubara frequency calculated within ED DCA for $U=1 - 4$ 
at $T=0.01$. 
}\label{Sij.ED}\end{figure}

\begin{figure}  [t!] 
\begin{center}
\includegraphics[width=5.0cm,height=8.5cm,angle=-90]{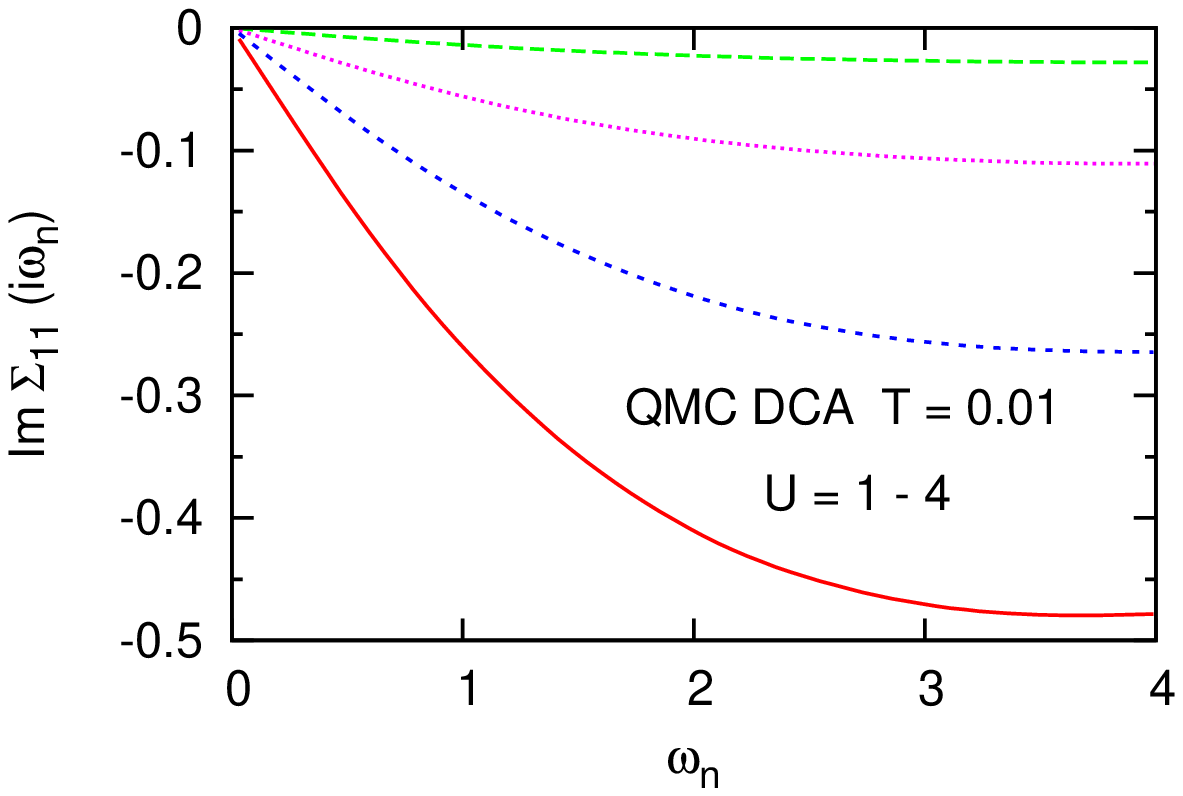}
\includegraphics[width=5.0cm,height=8.5cm,angle=-90]{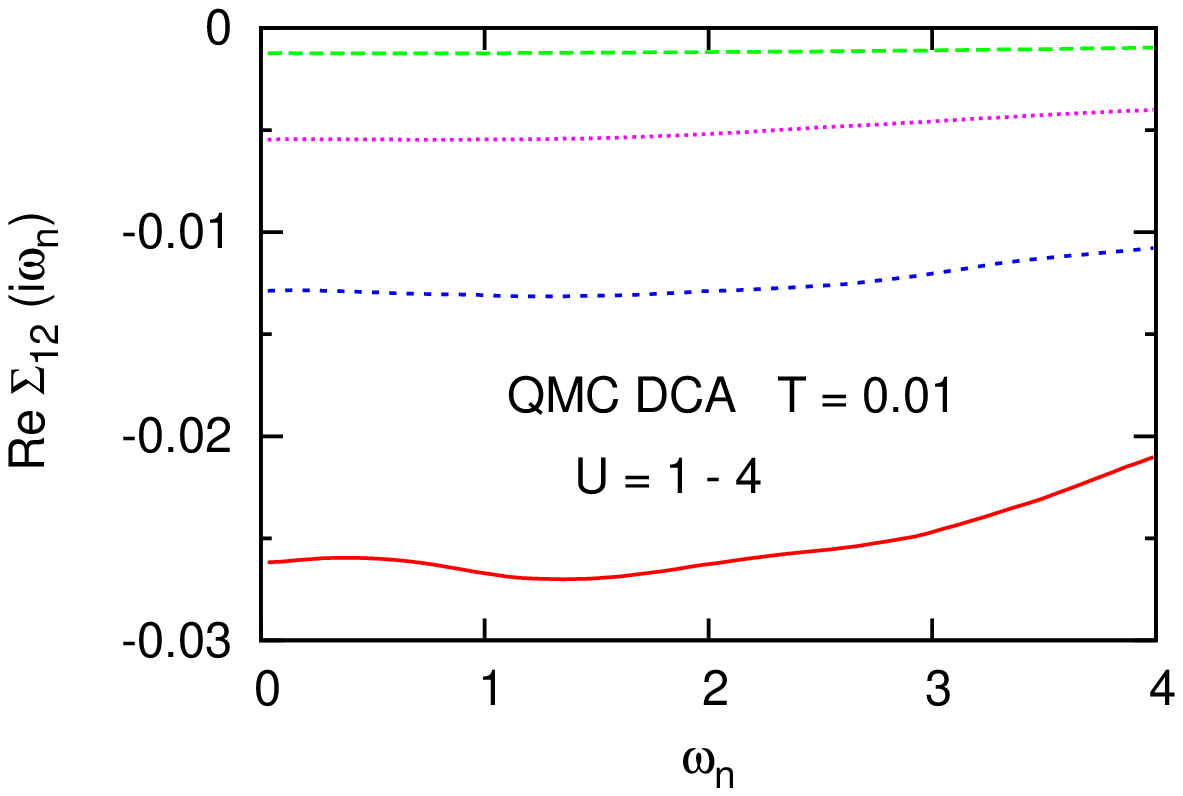}
\includegraphics[width=5.0cm,height=8.5cm,angle=-90]{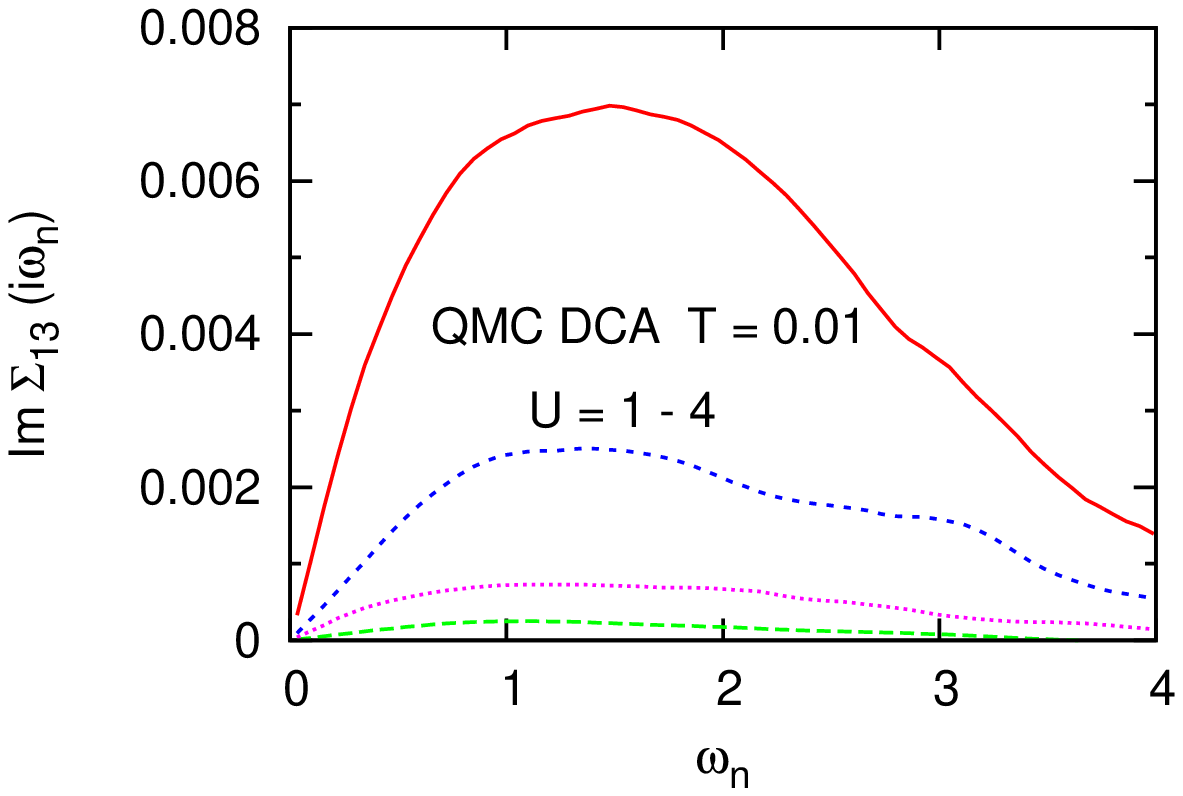}
\end{center}\vskip-3mm
\caption{(Color online)
Self-energy components $\Sigma_{1i}(i\omega_n)$ ($i=1,2,3$) of honeycomb lattice
as functions of Matsubara frequency calculated within CTQMC DCA for $U=1 - 4$ at 
$T=0.01$.
}\label{Sij.QMC}\end{figure}

To illustrate the effect of Coulomb correlations in more detail, we show in
Fig.~\ref{Sij.ED} the self-energy components in the site basis for several values 
of $U$.  The corresponding results obtained within CTQMC DCA are 
depicted in Fig.~\ref{Sij.QMC}. There is good overall correspondence between 
these two impurity solvers, except for slightly different magnitudes of the 
off-diagonal components. We note, however, that Re\,$\Sigma_{12}$ and Im\,$\Sigma_{13}$
are approximately one and two orders of magnitude smaller than Im\,$\Sigma_{11}$, 
respectively. As can be seen in Fig.~\ref{Gij}, these differences have only a minor 
effect on the variation of the Green's function components with increasing Coulomb
energy.

The crucial question in the case of the honeycomb lattice is how Coulomb correlations
influence the energy bands in the vicinity of the Dirac points. The self-energy 
at these points can be shown to have the simple form: 
\cite{PRB2011}
\begin{equation}
  \Sigma(K,i\omega_n) \approx i\omega_na + \frac{b^2}{i\omega_n(1-a)}, 
                   \hskip3mm  \omega_n\rightarrow0, 
\label{SK}
\end{equation}
where the coefficients are given by
\begin{eqnarray}
    a &=& {\rm Im}\,[\Sigma_{11}(i\omega_n)-\Sigma_{13}(i\omega_n)]/\omega_n \\
    b &=& {\rm Re}\,[\Sigma_{12}(i\omega_n)-\Sigma_{14}(i\omega_n)]
\end{eqnarray}
in the limit $\omega_n\rightarrow0$.
Thus, $\Sigma(K,i\omega_n)$ is imaginary as expected for particle-hole symmetry 
near the Dirac points. Moreover, this self-energy consists of metallic 
($\sim i\omega_n$) and insulating ($\sim 1/i\omega_n$) contributions, where the 
latter term is a direct consequence of the fact that $\Sigma_{12}\ne\Sigma_{14}$.
The presence of this term implies 
Re\,$\Sigma(K,\omega) \approx b^2/[\omega(1-a)]$ at real $\omega$. In the 
low-temperature limit, this expression yields an excitation gap of magnitude 
$\Delta \approx 2 \sqrt{\vert c \vert}$, where $c= b^2/(1-a)$. 
A similar insulating contribution to the self-energy was recently found in 
Ref.~\onlinecite{seki}. Presumably, this insulating term is also present in the 
ED calculations reported in Refs.~\onlinecite{helu} and \onlinecite{yu}. 
At finite $T$, the gap 
is smoothened out so that it becomes difficult to determine its boundaries. 
In contrast, as discussed in Section II, DCA preserves the bulk symmetry, 
so that $\Sigma_{12}=\Sigma_{14}$ and $\Delta=0$. 
Thus, the DCA self-energy at the Dirac points is purely metallic, where the 
increasing magnitude of the coefficient $a$ implies increasing quasi-particle 
broadening and shift of spectral weight towards the Fermi level as $U$ increases.
From the initial slope of Im\,$\Sigma_{11}$ at $U=4$ we obtain an effective mass 
enhancement of about $m^*/m\approx 1.3$.     

The above discussion demonstrates that the presence or absence of the insulating 
contribution to $\Sigma(K,i\omega_n)$ is not caused by the impurity solver used 
in the CDMFT or DCA calculations. In fact, the good agreement between ED and CTQMC, 
for both CDMFT and DCA, suggests that in the case of the honeycomb lattice one bath 
level per impurity orbital is sufficient for an accurate fit of the bath Green's 
function. The reason is that, because of the semi-metallic nature of the honeycomb 
lattice, the projection of the bath Green's function of the infinite lattice onto 
a finite-cluster Anderson Green's function is not plagued by the  
low-energy-low-temperature discrepancies that usually occur in the case of correlated 
metals. In these systems at least two bath levels per impurity orbital are 
typically required and very low temperatures must be avoided.\cite{EDR}

\section{Summary}

The role of Coulomb correlations in Hubbard model for the honeycomb lattice has 
been studied within finite-temperature exact diagonalization and continuous-time
quantum Monte Carlo combined with the dynamical cluster approximation. 
The unique feature of DCA is that it preserves the translation invariance so 
that the system at small values of $U$ is semi-metallic. In contrast, CDMFT 
violates translation symmetry which implies the opening of an excitation gap 
at arbitrarily small $U$, regardless of the impurity solver.  
This gap is therefore an artifact caused by the lack of long-range crystal 
symmetry and does not correspond to a true Mott gap. At larger values of $U$,
however, many-body interactions are dominated by short-range correlations and 
translation symmetry seizes to be important. DCA then becomes less accurate since 
it overemphasizes semi-metallic behavior. Thus, for $U \approx 5$, CDMFT is 
preferable and reveals a Mott gap in qualitative agreement with large-scale QMC 
calculations. 

In the case of the honeycomb lattice, DCA and CDMFT may therefore be viewed 
as complementary cluster approaches. As DCA preserves translation symmetry, 
it is more appropriate in the semi-metallic phase at small $U$ where long-range
order is a prerequisite for the description of the weakly correlated Dirac cones. 
The condition $\Sigma_{12}=\Sigma_{14}$ which guaranties this symmetry, however, 
is unrealistic at larger $U$, when short-range correlations within the six-site 
unit cell begin to dominate. Thus, in the region of the Mott phase, CDMFT is more 
suitable. As a result of these inherent limitations of both cluster schemes, the 
critical Coulomb interaction defining the precise boundary between these phases 
is at present difficult to determine within either CDMFT or DCA. 
We emphasize that this difficulty is not related to the finite size or symmetry 
of the bath used in ED. On the contrary, within CDMFT as well as DCA, the ED 
self-energies agree well with the corresponding CTQMC results. 

It is interesting to inquire why the remarkable difference between CDMFT and DCA 
for the honeycomb lattice discussed in this paper does not also manifest itself 
in other systems, such as the Hubbard models for square and triangular lattices. 
In these cases, long-range order is mainly responsible for the logarithmic 
divergence of the van Hove singularities of the density of states. 
Thus, any lack of perfect translation symmetry would give rise to a rounding 
of this peak, an effect that would be difficult to distinguish from broadening 
induced by finite temperature and quasi-particle damping. In contrast, any 
rounding of Dirac cones induces the opening of a gap. In this regard, 
the Dirac cones of the honeycomb lattice correspond to a rather peculiar special
situation that does not arise in most cases which have been studied previously 
within CDMFT and DCA at finite temperatures.

\bigskip

{\bf Acknowledgements}\ \ 
The ED DCA calculations were carried out on the J\"ulich Juropa machine.  
W. W. acknowledges support from the Natural Sciences and Engineering Research 
Council of Canada. A. L. likes to thank Profs. Lu and Seki for sending their
data shown in Fig.~\ref{gap}.


\end{document}